\documentclass[prl,twocolumn,showpacs,preprintnumbers,amsmath,amssymb,superscriptaddress]{revtex4-1}

\usepackage{graphicx}% Include figure files
\usepackage{epsfig}
\usepackage{amsmath}
\usepackage{amssymb}
\usepackage{textcomp}
\usepackage{dcolumn}% Align table columns on decimal point
\usepackage{bm}% bold math
\usepackage{braket}
\PassOptionsToPackage{numbers,sort&compress}{natbib}
\usepackage{pdfpages}%for attaching the supplemental

\makeatletter
\g@addto@macro\normalsize{%
  \setlength\abovedisplayskip{4pt}
  \setlength\belowdisplayskip{4pt}
  \setlength\abovedisplayshortskip{4pt}
  \setlength\belowdisplayshortskip{4pt}
}
\makeatother

\newcommand{\BibitemShut}[1]{} %handles an error due to an outdated natbib

\begin{document}

\newcommand*{\MAINZ}{QUANTUM, Institut f\"ur Physik,
Universit\"at Mainz, Staudingerweg 7, 55128 Mainz, Germany}
\affiliation{\MAINZ}
\newcommand*{\IQOQI}{Institut f\"ur Quantenoptik und Quanteninformation, \"Osterreichische Akademie der Wissenschaften, Technikerstra{\ss}e 21a, 6020 Innsbruck, Austria}
\newcommand*{\IBK}{Institut f\"ur Experimentalphysik, Universit\"at Innsbruck, Technikerstra{\ss}e 25, 6020 Innsbruck, Austria}
%\affiliation{\MAINZ}
%\affiliation{\IQOQI}
%\affiliation{\IBK}
\homepage{http://www.quantenbit.de}

\title{Measurement of dipole matrix elements with a single trapped ion }
\author{M. Hettrich}\email{hettrich@uni-mainz.de}\affiliation{\MAINZ}
\author{T. Ruster}\affiliation{\MAINZ}
\author{H. Kaufmann}\affiliation{\MAINZ}
\author{C. F. Roos}\affiliation{\IQOQI}\affiliation{\IBK}
\author{C. T. Schmiegelow}\affiliation{\MAINZ}
\author{F. Schmidt-Kaler}\affiliation{\MAINZ}
\author{U. G. Poschinger}\affiliation{\MAINZ}

\date{\today}% It is always \today, today,

\begin{abstract}
We demonstrate a method to determine %knapper, konziser schreiben: "We determine....?"
dipole matrix elements by comparing measurements of dispersive and absorptive light ion interactions. We measure the matrix element pertaining
to the Ca II H line, i.e. the 4$^2$S$_{1/2} \leftrightarrow $ 4$^2$P$_{1/2}$ transition of  $^{40}$Ca$^+$, for which we find the value 2.8928(43) ea$_0$.  Moreover, the method allows us
to deduce the lifetime of the 4$^2$P$_{1/2}$ state to be 6.904(26) ns, which is in agreement with predictions from recent theoretical calculations and resolves a longstanding discrepancy
between calculated values and experimental results.
\end{abstract}
%  37.10.Ty   Ion trapping
%  32.80.Qk   Coherent control of atomic interactions with photons
\maketitle
Methods for trapping and cooling single or few atoms, molecules or ions and manipulating them at the quantum level have opened up new avenues for precision laser spectroscopy. In particular,
quantum logic techniques \cite{PIETSCHMIDT2005} have enabled a new accuracy regime of timekeeping with optical atomic clocks \cite{ROSENBANDCLOCK2}. In contrast to atomic transition
frequencies, dipole matrix elements and radiative lifetimes are still notoriously hard to determine at high accuracy, but are important for the quantification of black body radiation shifts
of atomic clocks \cite{SAFRONOVABBR2011}, interpretation of astrophysical spectra \cite{RAUSCHER2006,CARLSSON2012}, novel
approaches for the search for physics beyond standard model \cite{FORTSON1993, KOERBER} and for testing the accuracy of atomic structure calculations \cite{SAFRONOVA2011}. 

Regarding measurements of radiative lifetimes and transition matrix elements, established methods e.g. based on ion beams have been successfully complemented by novel techniques based
on trapped particles. For $^{87}$Rb,  dipole matrix elements have been determined on the 10$^{-3}$ uncertainty level by diffraction in a condensate \cite{HEROLD2012}, while for the 6p${^2}$P$_{1/2}^{o}$
state of $^{174}$Yb$^+$, the radiative lifetime has been measured by time-resolved counting of photons emitted from a single trapped ion \cite{MONROEYB}. A related technique was used
for neutral $^{171}$Yb in an optical lattice \cite{BELOY2012}. 
\begin{figure}[t]\begin{center}
\includegraphics[width=0.45\textwidth]{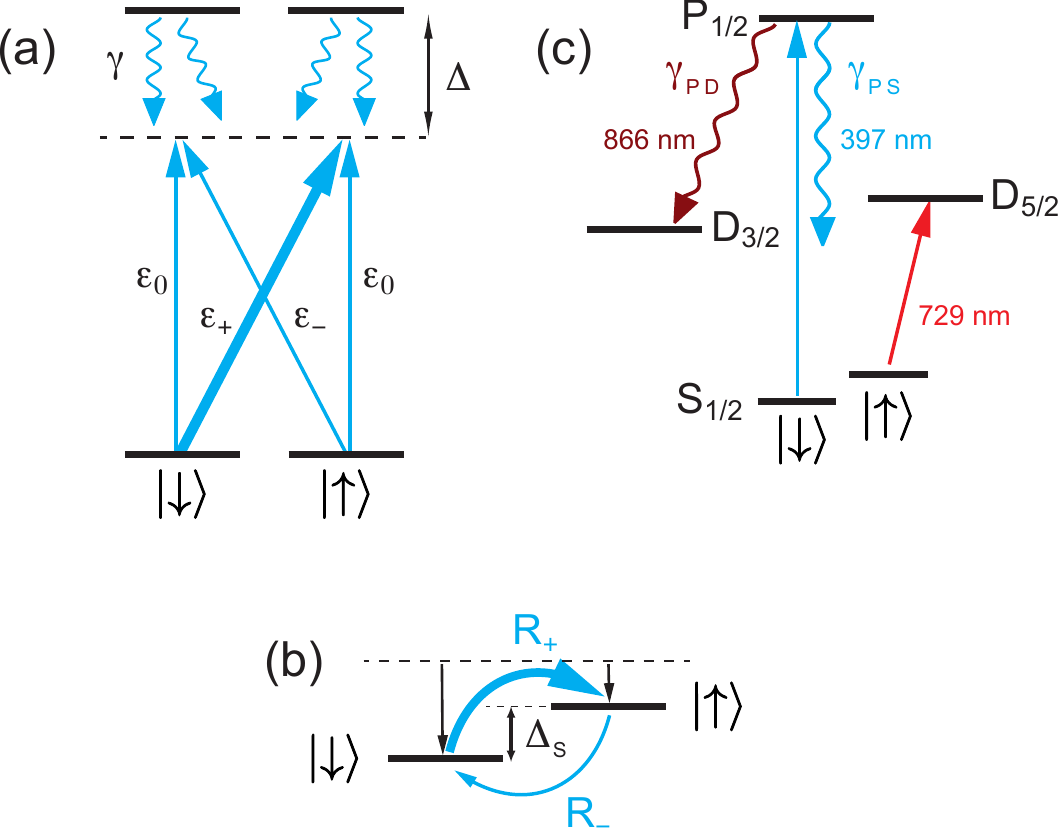}
\caption{ (color online) Measurement scheme: (a) a laser off-resonantly couples two Zeeman ground state levels to an excited state with decay rate $\gamma$. (b) The coupling gives
rise to a differential ac Stark shift $\Delta_S$ between the Zeeman levels and incoherent redistribution of population at rates $R_\pm$. (c) Relevant energy levels in $^{40}$Ca$^+$.
The shelving to the metastable 3$^2D_{5/2}$ state
as well the 3$^2D_{3/2}$ state acting as a population sink are indicated. State populations, which result in a bright (dark) detection event are marked by an open (closed) circle.}
\label{fig:levelscheme}
\end{center}
\end{figure}

The species Ca$^+$ is widely used in quantum optics experiments, and its II H line led to the discovery of the interstellar medium \cite{HARTMANN1904}. For $^{40}$Ca$^+$, branching
ratios between different decay channels have been determined at uncertainties approaching the 10$^{-5}$ level \cite{GERRITSMA,HARTMUT2013}, and lifetimes of metastable states have been
accurately measured \cite{BLATTLIFETIME}. 
The radiative lifetime of the 4$^2$P$_{1/2}$ excited state of $^{40}$Ca$^+$ has been determined to be 7.098(20)~ns \cite{JINCHURCH} by fluorescence measurements on a fast ion beam. However, this value disagrees with the most recent theoretical calculation \cite{SAFRONOVA2011} by more than 11 standard deviations, while similar calculations for alkaline-like species from Li to Fr, Mg$^+$, Ba$^+$ and Sr$^+$ are in good agreement with experimental results (see \cite{SAFRONOVA2011} and references therein).

In this work, we determine the radiative decay rates and the lifetime of the 4$^2$P$_{1/2}$ excited state of $^{40}$Ca$^+$ together with the dipole matrix element of the 4$^2$S$_{1/2} \leftrightarrow\,$4$^2$P$_{1/2}$ transition. The cornerstone of our scheme is the comparison between the dispersive and absorptive interactions, which occur upon driving this transition with an off-resonant laser, see Fig. \ref{fig:levelscheme}. As the method is based on the discrete discrimination of atomic states of a single trapped particle \cite{DIDIRMP,UNSERJPHYSB}, it is robust against many systematic error sources which affect other existing methods. 

The off-resonant laser is characterized by its detuning $\Delta$ from the 4$^2$S$_{1/2} \leftrightarrow\,$4$^2$P$_{1/2}$ transition, the Rabi frequency $\Omega$ and the relative amplitudes $\epsilon_q$, which characterize the circular ($q=\pm 1$, denoted '$\pm$' henceforth) and $\pi$ ($q=0$) polarization components.

The \textit{dispersive} interaction with the laser field, detuned by a frequency $\Delta$ from resonance, causes ac Stark shifts \cite{RUDI,WINELAND2007} of the energy levels. Specifically, we are interested in the differential ac Stark shift $\Delta_S$ between the two Zeeman-sublevels of the electronic ground state $\ket{S_{1/2},m_S=\pm 1/2}$, denoted henceforth as $\ket{\uparrow}$ and $\ket{\downarrow}$. It reads 
\begin{equation}
\Delta_S=\frac{1}{3}\frac{\Omega^2}{4\Delta}(\epsilon^2_{-}-\epsilon^2_{+}),
\label{eq:starkshift}
\end{equation}
and is obtained by a spin echo measurement technique \cite{HAEFFNER2003}.

The \textit{absorptive} interaction mediated by the same laser field manifests itself through inelastic Raman scattering, i.e. spin flips \cite{OZERI2005,UYS}.
We denote the spin flip rate from $\ket{\downarrow}$
to $\ket{\uparrow}$ with $R_+$ and the rate in the inverse direction respectively with $R_-$. For an optical field with arbitrary polarization they read \cite{WINELAND2007}
\begin{equation}
R_{\pm}=\gamma_{PS}\frac{\epsilon^2_{\pm}+\epsilon^2_{0}}{9}\frac{\Omega^2}{4\Delta^2},
\label{eq:scatterrates}
\end{equation}
where $\gamma_{PS}$ is the radiative decay rate from the  P$_{1/2}$ to the S$_{1/2}$ state. 
The spin flip rates $R_{\pm}$ are experimentally determined by monitoring the population in $\ket{\uparrow}$ and $\ket{\downarrow}$ during the interaction.
Note that the $\pi$-polarized field component drives the two spin-flip directions equally strong, whereas each of the the circular components drives only one pathway, corresponding to optical
pumping in the limit of purely circularly polarized field component. The occurrence of elastic (Rayleigh) scattering is taken into account by the prefactor. The detuning $\Delta$ is chosen such that the conditions $\vert\Delta\vert\gg\Omega,\gamma_{PS}$ for  Eqs. \ref{eq:starkshift} and \ref{eq:scatterrates} to hold are met.

By combining Eqs. \ref{eq:starkshift} and \ref{eq:scatterrates}, we obtain the decay rate $\gamma_{PS}$:
\begin{equation}
\gamma_{PS}=3\Delta\frac{\delta R}{\Delta_S},
\label{eq:central}
\end{equation}
with $\delta R=R_--R_+$. All quantities on the rhs of \ref{eq:central} are experimentally accessible.

As both interactions are driven with the same optical field, the Rabi frequency $\Omega$ and the polarization amplitudes $\epsilon_q$ cancel out in Eq. \ref{eq:central} and notorious error sources are eliminated. These quantities are given by the electric field amplitude of the off-resonant laser at the position of the ion, which reads $\mathbf{E}=\sum_{q=\pm,1}\vert \mathbf{E}\vert\epsilon_q \mathbf{e}_q+c.c.$. Here, $ \mathbf{e}_q$ are the spherical basis vectors. The  polarization amplitude $\epsilon_q$ define the coupling between Zeeman sublevels with $m_P-m_S=q$ with effective Rabi frequencies scaled with the respective Wigner 3j-symbol, $\Omega \epsilon_q \left( \begin{smallmatrix} 1/2 & 1 & 1/2 \\ -m_S & q & m_P \end{smallmatrix} \right)$. Here $\Omega=2\vert\mathbf{E}\vert\mathcal{D}/\hbar$ is defined as the base Rabi frequency, with the reduced dipole matrix element $\mathcal{D}=\langle S_{1/2} \| \mathbf{\hat{d}} \| P_{1/2}\rangle$. 

We  take an additional decay channel into account, for the case of $^{40}$Ca$^+$ decay from the 4$^2$P$_{1/2}$ to the 3$^2$D$_{3/2}$ state occurs at rate $\gamma_{PD}$.  For that, we assume that only circular components are present in the beam, i.e. $\epsilon_0^2
\approx 0$. The rates at which population from $\ket{\uparrow} (\ket{\downarrow})$ is
sunk in the metastable 3$^2$D$_{3/2}$ state are then proportional to $R_{\pm}$: 
\begin{equation}
R_{\uparrow(\downarrow) D}=3(\gamma_{PD}/\gamma_{PS})R_{-(+)}\equiv b R_{-(+)},
\label{eq:scatterratesD}
\end{equation}
with the leak factor $b$, which is also accessible from the measured time-dependent populations. Together with $\gamma_{PD}$, we obtain also the 4$^2$P$_{1/2}$ state's lifetime $\tau$. Finally, using
$\gamma_{PS}$ and the resonance wavelength $\lambda_{PS}$ of the 4$^2$S$_{1/2} \leftrightarrow\,$4$^2$P$_{1/2}$ transition, we can deduce the reduced matrix element $\mathcal{D}$: 
\begin{equation}
\mathcal{D}^2= 2 \gamma_{PS} \frac{3\epsilon_0 \hbar}{8 \pi^2}\lambda_{PS}^3.
\label{eq:matrixElement}
\end{equation}

We perform measurements on a single $^{40}$Ca$^+$ ion stored in a segmented Paul trap \cite{SCHULZ2008}. The relevant energy levels and transitions are depicted in Fig. \ref{fig:levelscheme}. 
A magnetic field splits the Zeeman sublevels of the
electronic ground state $\ket{\uparrow}$ and $\ket{\downarrow}$ by $2\pi\times$13.7~MHz. The ion is Doppler cooled on the 4$^2$S$_{1/2} \leftrightarrow $4$^2$P$_{1/2}$ (cycling) transition near 397~nm and
afterwards prepared in either $\ket{\uparrow}$ or $\ket{\downarrow}$ by optical pumping. Now, the ion is illuminated with
light near 397~nm, detuned by $\Delta$ from the  4$^2$S$_{1/2} \leftrightarrow\,$4$^2$P$_{1/2}$ transition to induce both the dispersive and absorptive interactions, which allow for determining $\Delta_S$ and $R_{\pm}$.
Spin read-out is accomplished by shelving population from the $\ket{\uparrow}$ level to the metastable 3$^2$D$_{5/2}$ state by means of rapid adiabatic passage (RAP) pulses \cite{WUNDERLICH,UNSERJPHYSB}. This allows
for discrimination between $\ket{\uparrow}$ and $\ket{\downarrow}$ as the linewidth of the $4^2S_{1/2} \leftrightarrow 3^2D_{5/2}$ quadrupole transition and the bandwidth of the RAP pulses are much smaller than the Zeeman
splitting. Then, the ion is illuminated by laser fields driving the cycling transition and the 3$^2$D$_{3/2}\leftrightarrow$4$^2$P$_{1/2}$ transition near 866~nm,
such that fluorescence is detected on a photo-multiplier tube if the ion has not been shelved. Conversely, the probability to not detect fluorescence, i.e. for a dark event, corresponds to the probability that the ion has been shelved from $\ket{\uparrow}$ to the metastable state.

 The off-resonant laser light is provided by an amplified and frequency-doubled diode laser system, stabilized in both wavelength
($\delta \Delta/\Delta \lesssim 0.5\cdot 10^{-3}$) and intensity ($\delta I/I \lesssim 0.5\cdot 10^{-3}$). The beam is aligned along the quantizing magnetic field and predominantly $\sigma^+$ polarized, such that
$R_+ \gg R_-$, and $\epsilon_0^2\approx 0$, justifying the approximation for Eq. \ref{eq:scatterratesD}.

\begin{figure}[htp]\begin{center}
\includegraphics[width=0.48\textwidth]{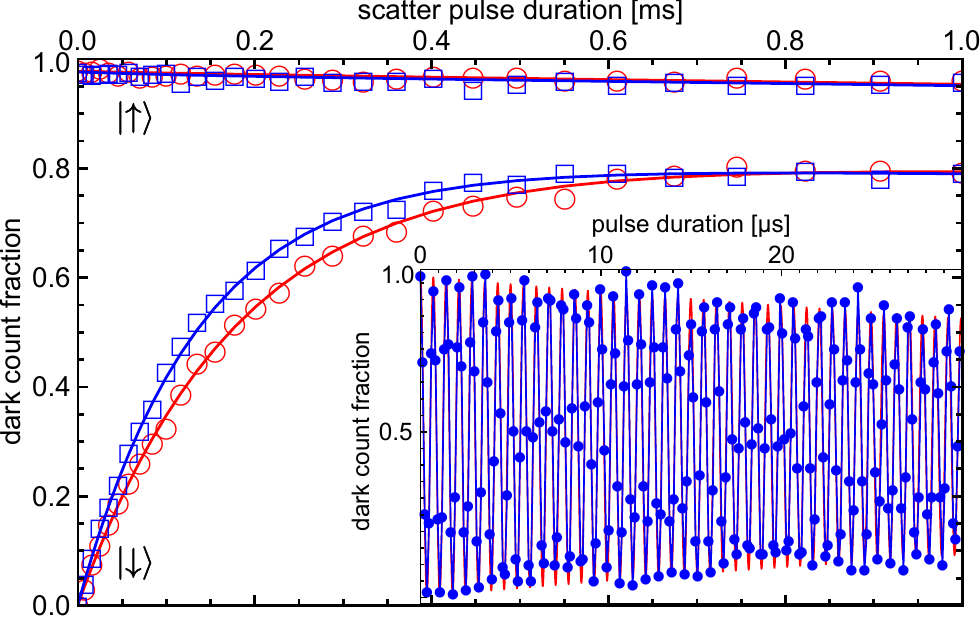}
\caption{ (color online) Dark event fraction versus time of exposure to an off-resonant laser pulse for initializations in $\ket{\uparrow}$ and $\ket{\downarrow}$. Data points indicated
by blue squares (red circles) are measured with the laser detuned by 12.03~GHz (13.94~GHz) from resonance. Solid lines show the fits to the model  Eq. (\ref{eq:scattersol}). The inset
shows the dark event fraction versus pulse duration for a spin-echo experiment. The blue line connects the datapoints, the red line is the resulting fit. The oscillation frequency directly
corresponds to the
differential ac Stark shift $\Delta_S$.}
 \label{fig:scatter}
\end{center}\end{figure}

The ac Stark shift $\Delta_S$ is measured with a spin-echo sequence, where
a  $\pi$/2 pulse on the stimulated Raman transition between $\ket{\uparrow}$ and
$\ket{\downarrow}$ is followed by  a $\pi$ pulse, and another $\pi$/2 pulse concludes the sequence. The delay time between the pulses is constant. During the first delay, the ion is exposed to a square-pulse
of variable duration from the off-resonant laser. We probe 250 different shift pulse times, each with 150 interrogations. The pulse durations are spaced by 120~ns, such that a signal with
about 40 oscillation periods is obtained. The differential ac Stark shift is the frequency of this oscillatory dark event fraction (see inset of Fig. \ref{fig:scatter}).

For the measurement of the spin-flip rates $R_{\pm}$, the ion is exposed to pulses of variable duration up to 1~ms of the off-resonant laser after preparation. This changes the spin populations as depicted in Fig.  \ref{fig:scatter}. The time gap between initialization and readout is
kept constant irrespectively of the scatter pulse duration to avoid systematic effects.

The spin-flip dynamics can be described by the rate equations:
\begin{eqnarray}
\dot{p}_{\uparrow}&=&-R_-(1+b)\;p_{\uparrow}+R_+p_{\downarrow} \nonumber \\
\dot{p}_{\downarrow}&=&-R_+(1+b)\;p_{\downarrow}+R_-p_{\uparrow}.
\label{eq:scatterdyn}
\end{eqnarray}
The solution of Eqs. \ref{eq:scatterdyn} is
\begin{eqnarray}
p_{\uparrow}^{(\uparrow)}(t)&=& \tfrac{1}{\tilde{R}}e^{-\tfrac{1}{2}\bar{R}t}\left(\tilde{R} \cosh(\tfrac{\tilde{R}t}{2})-(1+b)\delta R\sinh(\tfrac{\tilde{R}t}{2})\right)\nonumber \\
p_{\uparrow}^{(\downarrow)}(t)&=&\tfrac{2}{\tilde{R}} e^{-\tfrac{1}{2}\bar{R}t}R_+\sinh(\tilde{R}t/2),
\label{eq:scattersol}
\end{eqnarray}
with $p_{\uparrow}^{(\uparrow)}$ corresponding to initialization in $\ket{\uparrow}, p_{\uparrow}(t=0)=1$, and $p_{\uparrow}^{(\downarrow)}$ for initialization in $\ket{\downarrow}, p_{\downarrow}(t=0)=1$. We use $\bar{R}=(1+b)(R_-+R_+)$ and $\tilde{R}^2=\bar{R}^2-4 b(2+b) R_-R_+$. 

\begin{figure}[htp]\begin{center}
\includegraphics[width=0.48\textwidth]{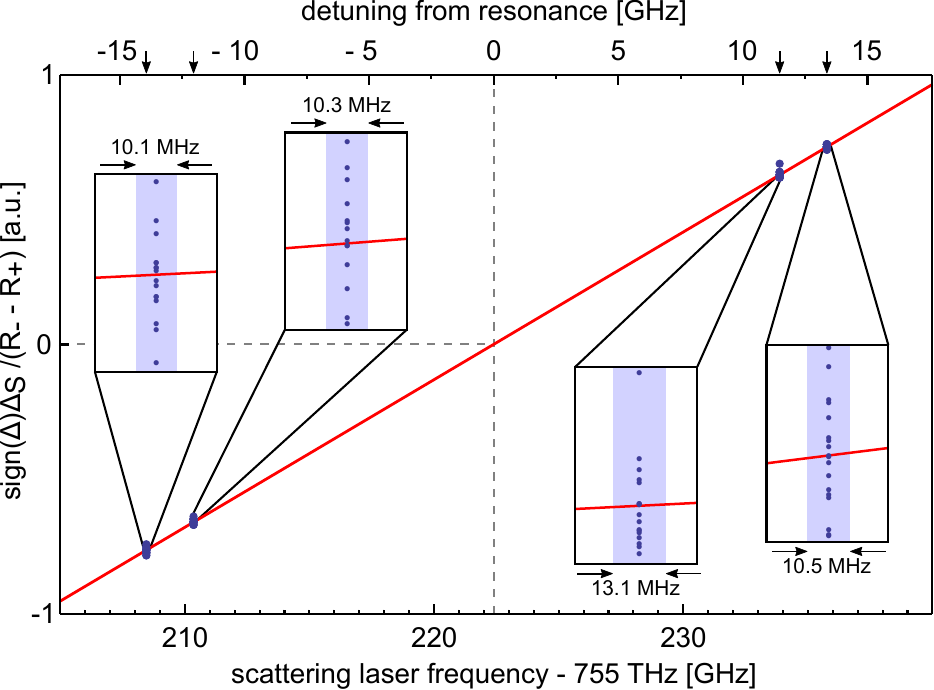}
\caption{(color online) Determination of the four detunings chosen for our measurements. For each of the 65 acquired datasets, we plot the quantity 
$\text{sign}(\Delta)\Delta_S/\delta R\propto\Delta$ versus the optical frequency as measured by a wavelength meter. The zero crossing of a linear fit,
 depicted by the solid red line, reveals the resonance frequency with a standard measurement uncertainty of $2\pi\times$21~MHz. The insets show details of the datasets measured at the different detunings
with the abscissa magnified 166 times. The shaded backgrounds indicate their uncertainties along the frequency axis. The arrows on the top frequency scale indicate the values of the
four different detunings at -13.94, -12.03, 11.52 and 13.42 GHz.}
 \label{fig:resonance}
\end{center}\end{figure}

For each measurement, the dark event fraction is determined by probing the ion 2500 times for 30 fixed scatter pulse durations.
Note that while the ac-Stark shift is given by the frequency of an oscillatory signal, the spin-flip rates are given by time-constants of exponentially decaying signals.
Thus, the latter measurement requires significantly more data to attain the same level of precision as the former. 
We extract the values for the differential scattering rate $\delta R$ and the leak factor $b$ 
by means of a Markov chain Monte Carlo parameter estimation, taking into account the binomial statistics of the shot noise in the spin readout. State preparation and measurement errors are taken into account by modeling the measured dark event fractions as a linear transformation of the values $p_{\uparrow}^{(\uparrow)}(t),p_{\uparrow}^{(\downarrow)}(t)$ from Eqs. \ref{eq:scattersol}.

We repeated ac Stark shift measurements and spin flip rate measurements in an interleaved fashion in order to capture drifts of the laser intensity and thus the Rabi
frequency $\Omega$. One measurement run consists of one spin flip rate measurement for preparation both in $\ket{\uparrow}$ and $\ket{\downarrow}$, preceded and followed by one ac Stark shift measurement.  In total, 65 of such measurement runs were performed, at a total acquisition time of about 50~h.
Four different values of the detuning $\Delta$ were used to show that no systematic effects with respect to this parameter are present. 
\begin{figure}[htp!]
\begin{center}
\includegraphics[width=0.5\textwidth]{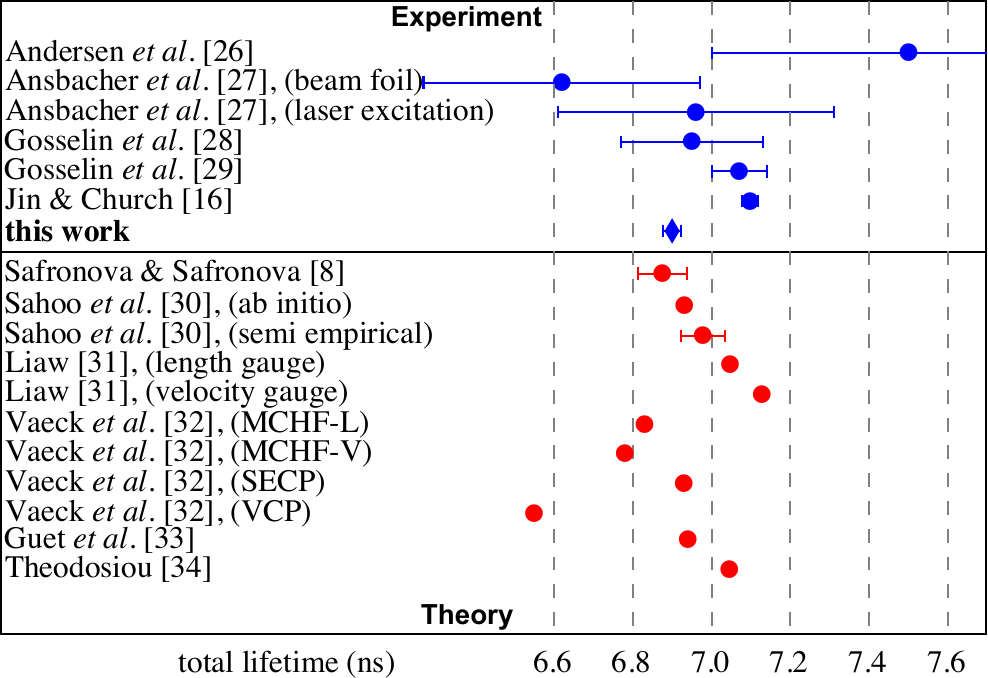}
\caption{Comparison of the measured lifetime with theoretical and experimental results for P$_{1/2}$ - level lifetimes from references \cite{ANDERSEN1970,ANSBACHER1985,GOSSELIN1988NIMB,
GOSSELIN1988PRA, JINCHURCH,SAFRONOVA2011,SAHOO2009,LIAW,VAECK,GUET,THEODOSIOU}}
\label{fig:resultsCompare}
\end{center}
\end{figure}
To determine the value of these detunings, we read the optical frequency from a
commercial wavelength meter (High
Finesse WSU 267) with better than 10~MHz precision. We perform a linear regression of the measured $\Delta_S$ divided by $\delta R$ versus the corresponding optical frequencies to obtain the resonance frequency, as
shown in Fig. \ref{fig:resonance}. Compared to direct fluorescence spectroscopy, this technique mitigates effects like power broadening, Zeeman splitting or micromotion induced broadening, which would make a determination
of the resonance frequency less accurate.

Finally, we combine the measured quantities  $\Delta, \Delta_S$ and $\delta R$, using Eq. \ref{eq:central}, to obtain the desired value $\gamma_{PS}$. Together with $b$ this also yields $\gamma_{PD}=\tfrac{b}{3}\gamma_{PS}$, and
the radiative lifetime is given by $1/\tau=\gamma_{PS}(1+\tfrac{b}{3})$.

Averaging over the data sets and taking into account the corrections discussed below, we obtain the resulting
values: $\gamma_{PS}=2 \pi \times 21.57(8)$~MHz and a 4$^2$P$_{1/2}\rightarrow$4$^2$S$_{1/2}$ branching fraction of $1/\left(1+\tfrac{b}{3}\right)= 0.93572(25)$. This leads to a lifetime
of $\tau=6.904(26)$~ns and a value of $\gamma_{PD}=2\pi\times 1.482(8)$~MHz. Using Eq. \ref{eq:matrixElement}, we obtain a value for the reduced dipole matrix element of $\mathcal{D}=2.8928(43)$~ea$_0$.

Experimental imperfections and model approximations lead to uncertainties and corrections for the value of $\gamma_{PS}$. These are summarized in Table \ref{tbl:ErrCorr}: Beyond the approximation in Eq. \ref{eq:scatterratesD}, a
possible $\epsilon_0$ polarization component has been taken into account, and yields corrections and uncertainties on the $10^{-3}$ level. 
The uncertainty of the detuning $\Delta$ is determined by the finite precision of the wavelength meter, and by the uncertainty of the determination of the resonance frequency, see Fig. \ref{fig:resonance}. The random measurement errors for $\delta R$, determined by the amount of acquired data, is another significant contribution to the uncertainty budget.
\begin{table}[htp!]
\begin{tabular}{p{5cm} p{1.7cm}p{1.7cm}}\hline\hline
\textbf{Effect} & \textbf{Shift}$\times 10^{-3}$ & \textbf{Unc.}$\times 10^{-3}$  \\\hline
Residual $\epsilon_0$ polarized light & 3.3 &  2.9\\
Uncert. of resonance frequency & - & 1.6 \\
Stat. uncertainty of $\delta R$ & - & 1.5 \\
Residual near-resonant light & -0.4 & 0.5\\
Wavemeter precision & - & 0.4 \\
$D_{3/2}$ depletion & 0.3 & 0.2 \\
Uncertainty of $\Delta_S$ & - & 0.2\\
Influence of micro motion & - & 0.1 \\
Influence of P$_{3/2}$ state & 0.3 & $<$ 0.1\\
Residual line-broadening effects & 0.1 & $<$ 0.1 \\
Total & +3.6& 3.7 \\\hline\hline
\end{tabular}
\caption{List of relative corrections and measurement uncertainties for $\gamma_{PS}$. The specified values result from averaging over the 65 sets of measurement data. They are added in quadrature to obtain the resulting final uncertainty.
\vspace{6pt}}
\label{tbl:ErrCorr}
\end{table}
Residual resonant light close to 397~nm resulting from imperfect laser switch-off causes relative corrections and uncertainties in the  $10^{-4}$ regime. Beyond the rate equation
model Eq. \ref{eq:scatterdyn}, we include a correction for the finite lifetime of the 3$^2$D$_{3/2}$ state or its depletion by residual light near 866~nm. $\Delta_S$ can be measured
by orders of magnitude more precisely than $\delta R$ and $b$, however the precision is limited by drift effects, quantified by monitoring the ac Stark shift over the entire data acquisition
time. Excess micromotion of the ion also causes a small systematic uncertainty via frequency modulation of the off-resonant light.  Furthermore, we include corrections due to the presence
of the 4${^2}$P$_{3/2}$ state about $2\pi\times$~6.69~THz above the 4${^2}$P$_{1/2}$ state, and due to the power-broadened Lorentzian lineshape beyond the assumption $\vert\Delta\vert\gg\Omega,\gamma_{PS}$.
A detailed discussion of corrections and uncertainties is presented in the supplemental material \cite{scatter_supplemental}.

Our value for $\tau$ is compared to previously reported values in Fig. \ref{fig:resultsCompare}. We find that our value agrees with the latest theory predictions \cite{SAFRONOVA2011},
while it is in substantial  disagreement with the most recent experimental result \cite{JINCHURCH}. Our value for the branching fraction is in agreement with the results from recent
measurements \cite{HARTMUT2013}. To our knowledge, there is no experimentally determined value of $\mathcal{D}$ so far. The value reported in our work is in agreement with the calculation from \cite{SAFRONOVA2011}.

Furthermore, using the measured value for $\mathcal{D}$, we infer the reduced matrix element pertaining to the 4$^2$S$_{1/2}\leftrightarrow$4$^2$P$_{3/2}$ transition,
which results to  $\langle S_{1/2} \| \mathbf{\hat{d}} \| P_{3/2}\rangle =\sqrt{2}\mathcal{D}=$~4.091(6)~ea$_0$.
Both reduced dipole matrix elements enter in the calculation of the blackbody-radiation (BBR) shift of the 4$^2$S$_{1/2}\leftrightarrow$3$^2$D$_{3/2}$ and 4$^2$S$_{1/2}\leftrightarrow$3$^2$D$_{5/2}$
quadrupole transitions, which are widely used testbeds for high precision laser spectroscopy \cite{CHWALLA2009}. The BBR shift is among the  major contributions to the uncertainties of clock transition frequencies, for species such as $^{43}$Ca$^+$ \cite{ARORA2007}, Sr$^+$\cite{Dube2013}, Sr \cite{SAFRONOVA2013,Ye2014} and Yb \cite{Ludlow2015}. These  are used for state-of-the art optical clock experiments, and some of them are discussed as new SI frequency standards. Our result adds to existing work \cite{Beloy2014,Dube2014,Middelmann2012} validating computational methods used to predict the
BBR shift in optical clocks.

In summary, we demonstrate a novel method for the measurement of dipole matrix elements, which works despite the presence of additional decay channels. We attain an uncertainty on the 10$^{-3}$ level. All major error sources can be potentially mitigated, such that measurements of radiative decay rates at unprecedentedly low uncertainties in the 10$^{-4}$ regime appear within reach with current technology. Our method is applicable to atom and ion species which allow for preparation and
readout of Zeeman or hyperfine sublevels.

\begin{acknowledgments}
We acknowledge financial support by the European commission within the IP SIQS and by the Bundesministerium f\"ur Bildung und Forschung via IKT 2020 (Q.com). UGP acknowledges funding by the Johannes-Gutenberg Universit\"at Mainz via internal university research funding grant \textit{Trapped ions in phase-stabilized standing waves}. CTS acknowledges support from the BMBF via the Alexander von Humboldt Foundation. We thank Vladan Vuletic, Rene Gerritsma and Marianna Safronova for helpful discussions.
\end{acknowledgments}

\bibliographystyle{apsrev4-1}
\bibliography{scatter_bib}

%the following is a hack to attach the supplemental material
%includepdf for the whole, doc at once doesn't work, so
%we do it like this
\newpage
$\,$
\includepdf[pages=1]{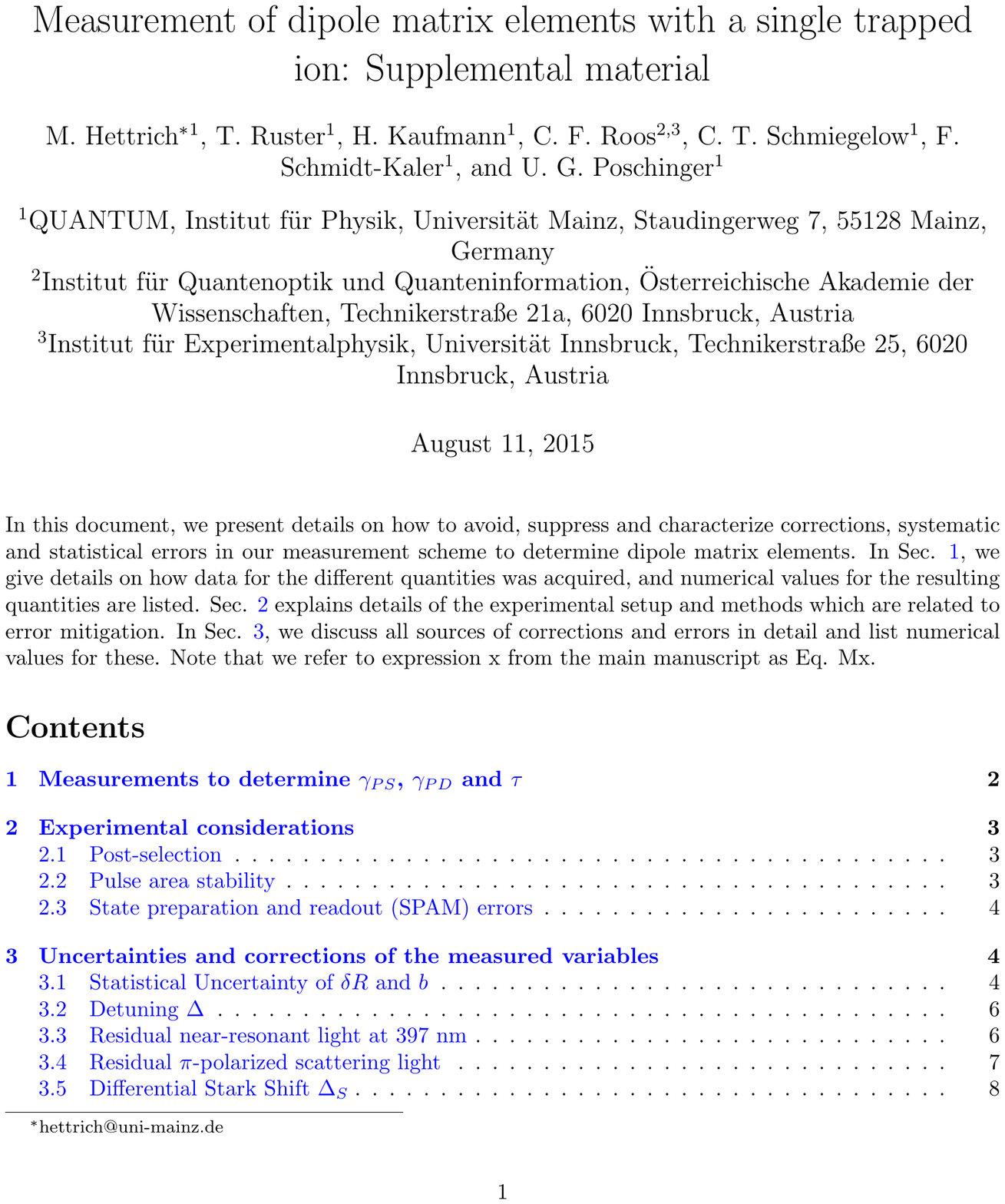}
$\,$
\includepdf[pages=2]{supplemental.pdf}
$\,$
\includepdf[pages=3]{supplemental.pdf}
$\,$
\includepdf[pages=4]{supplemental.pdf}
$\,$
\includepdf[pages=5]{supplemental.pdf}
$\,$
\includepdf[pages=6]{supplemental.pdf}
$\,$
\includepdf[pages=7]{supplemental.pdf}
$\,$
\includepdf[pages=8]{supplemental.pdf}
$\,$
\includepdf[pages=9]{supplemental.pdf}
$\,$
\includepdf[pages=10]{supplemental.pdf}
$\,$
\includepdf[pages=11]{supplemental.pdf}
$\,$
\includepdf[pages=12]{supplemental.pdf}
$\,$
\includepdf[pages=13]{supplemental.pdf}
$\,$


\section*{Supplementary material}

# Measurement of dipole matrix elements with a single trapped ion: Supplemental material

M. Hettrich[*1], T. Ruster[1], H. Kaufmann[1], C. F. Roos[2,3], C. T. Schmiegelow[1], F. Schmidt-Kaler[1], and U. G. Poschinger[1]

[1]QUANTUM, Institut für Physik, Universität Mainz, Staudingerweg 7, 55128 Mainz, Germany
[2]Institut für Quantenoptik und Quanteninformation, Österreichische Akademie der Wissenschaften, Technikerstraße 21a, 6020 Innsbruck, Austria
[3]Institut für Experimentalphysik, Universität Innsbruck, Technikerstraße 25, 6020 Innsbruck, Austria

August 11, 2015

In this document, we present details on how to avoid, suppress and characterize corrections, systematic and statistical errors in our measurement scheme to determine dipole matrix elements. In Sec. 1, we give details on how data for the different quantities was acquired, and numerical values for the resulting quantities are listed. Sec. 2 explains details of the experimental setup and methods which are related to error mitigation. In Sec. 3, we discuss all sources of corrections and errors in detail and list numerical values for these. Note that we refer to expression x from the main manuscript as Eq. Mx.

# Contents

*hettrich@uni-mainz.de

# 1 Measurements to determine $\gamma_{PS}$, $\gamma_{PD}$ and $\tau$

According to Eq. M3, the natural linewidth $\gamma_{PS}$ is determined by the detuning from resonance $\Delta$, the differential ac Stark shift $\Delta_S$ and the differential scattering rate $\delta R$. These quantities have been measured as follows:

- Detuning $\Delta$:

  The detuning $\Delta_R$ is determined by measuring the the absolute frequency $f_j$ of the scattering light for each detuning $j$. The number of detunings $D = 4$ (two on the blue side and two on the red side of the resonance) and the resonance frequency $f_{res}$ of the $S - P$-transition, which is determined as described in section 3.2.

- Scattering rates $R_+$ and $R_-$

  We have acquired a total number of 65 scattering curve pairs, distributed among four detunings $\Delta_i$, such that for each detuning $N_i = \{18, 17, 14, 16\}$ measurements were done. Each pair consists of one curve for initialization in $|\uparrow\rangle$, and one in $|\downarrow\rangle$. For each of these data sets, 30 different pulse times when the off-resonant beam is switched on were probed, ranging between 0 ms and 1 ms. The pulse times are unevenly distributed along the time axis to provide a more dense sampling for short pulse times. The 30 pulse times are probed in chunks of 25 repetitions, in a random permutation to avoid systematic effects. This is repeated 100 times, such that in total 2500 interrogations per pulse time are performed. For zero pulse time, in total 5000 interrogations are performed, which allows for a better accuracy in the determination of the readout parameters, c.f. Sec. 2.3. The measurements for each initialization level are performed within one sequence.

  The polarization of the off-resonant light is adjusted to be predominantly $\sigma^+$, i.e. $R_+ \gg R_-$, and it is not changed throughout all measurements. The two curves of each curve pair are *jointly* fitted to the model function. The fit provides the values $R_{+,i,j}$ and $R_{-,i,j}$ for curve pair $(i, j)$. $j$ identifies the detuning, and $i$ identifies the number of measurement for that respective detuning.

- Differential Stark shift $\Delta_S$

  For each scattering curve pair $(i, j)$, right before and right after the scattering measurement, Stark shift measurements are carried out, the mean of which, $\Delta_{S,i,j}$, is considered to be the Stark shift belonging to measurement $(i, j)$.

For each scattering curve pair, $\gamma_{PS,i,j}$ is calculated. We take the mean of those, and arrive at:

$$\gamma'_{PS} = 2\pi \times 21.48 \, \text{MHz} \tag{1}$$

2

Taking into account all corrections from section 3, that value reads:

$$\gamma_{PS} = 2\pi \times 21.57\,\text{MHz} \tag{2}$$

For each measured scattering curves $(i,j)$, the branching parameter $b_{i,j}$ is also given by the same fit mentioned above. With the mean of all $b_{i,j}$, $b$, the natural linewidths $\gamma_{PS}$ and $\gamma_{PD}$ are related as follows:

$$\gamma_{PD} = \frac{b}{3}\gamma_{PS} = 2\pi \times 1.48\,\text{MHz} \tag{3}$$

The total lifetime of the $P_{1/2}$-state $\tau$ is finally:

$$\tau = \frac{1}{\gamma_{PS} + \gamma_{PD}} = 6.90\,\text{ns} \tag{4}$$

The resulting values for these quantities are shown in table 1. The numbers are given for the four detunings $\Delta_i$, each averaged over the $N_i$ respective datasets.

| $i$ | $\Delta_i/2\pi$ (GHz) | $R_{i,+}\,(s^{-1})$ | $R_{i,-}\,(s^{-1})$ | $\Delta_{S,i}/2\pi$ (MHz) | $b_i$ | $\gamma_{PS,i}/2\pi$ (MHz) |
|---|---|---|---|---|---|---|
| 1 | -13.94 | 4474(78) | 47(7) | 1.3753 | 0.2063(48) | 21.45(33) |
| 2 | -12.03 | 6011(80) | 63(11) | 1.58819 | 0.2080(53) | 21.52(30) |
| 3 | 11.52 | 6514(125) | 101(7) | 1.64198 | 0.2056(45) | 21.49(39) |
| 4 | 13.42 | 4762(40) | 76(8) | 1.39819 | 0.2067(52) | 21.48(17) |

## 2 Experimental considerations

### 2.1 Post-selection

The data is post-selected in such a way that chunks of 25 interrogations where the ion is not found sufficiently bright after reset - indicating an impairment of Doppler cooling - are discarded. Similarly, chunks where the shelving was impaired were discarded, which can be identified by spuriously low dark count fractions. In total, less than 1% of all raw datapoints have been removed by this post-selection scheme.

### 2.2 Pulse area stability

Our measurement scheme relies on a constant optical power and polarization from the off-resonant laser near 397 nm at the location of the trapped ion. Several measures are employed to guarantee the stability. The laser light derived from a doubled, amplified laser system (Toptica TA-SHGpro) is coupled into a single mode fiber. At the fiber output, a polarizer filters undesired polarization components. After the polarizer, a small fraction of the optical power is picked and measured on a photodetector. The signal is used by a commercial intensity stabilization unit (NoiseEater, TEM Messtechnik), which acts on an electro-optical modulator before the fiber input. This way, intensity *and* polarization after the fiber output are actively stabilized. For switching, we employ an acousto-optical modulator (AOM) [1], which exhibits no measureable [2] beam pointing drift when switching between no power and maximum rf-power

---

[1] I-M110-3C10BB-3-GH27, Gooch&Housego

[2] Measured with a position sensitive detector PDP90A, Thorlabs

(2 W). The rf amplifier supplying the AOM is mounted on a temperature-stabilized water-cooled plate to avoid temperature-change-induced power drifts. The laser beam is focused such that zero and higher diffraction orders coincide at the position of the trapped ion, i.e. residual AOM induced beam pointing fluctuations are mitigated. Further mitigation is ensured by a relatively large beam waist of about 30 $\mu$m, and a careful positioning of the ion at the waist center by maximizing the ac Stark shift.
Diffraction angle and efficiency of an AOM weakly depends its temperature, and thus on the time-averaged rf-power. This is kept constant by fixing all pulse sequences for measurement of the ac Stark shifts and scattering rates to the same duration, and inserting additional rf pulses after readout to keep the average switched-on time constant.

## 2.3 State preparation and readout (SPAM) errors

Taking into account imperfect initialization, the experimental signals for the scattering rate measurements are modeled by a linear transformation of the populations from Eq. M6:

$$\begin{aligned} p_\uparrow^{(\uparrow)}(t) &= \epsilon_\downarrow + (1-\epsilon_\uparrow)\left(a_\uparrow \tilde{p}_\uparrow^{(\uparrow)}(t) + (1-a_\uparrow)\tilde{p}_\uparrow^{(\downarrow)}(t)\right) \\ p_\uparrow^{(\downarrow)}(t) &= \epsilon_\downarrow + (1-\epsilon_\uparrow)\left(a_\downarrow \tilde{p}_\uparrow^{(\downarrow)}(t) + (1-a_\downarrow)\tilde{p}_\uparrow^{(\uparrow)}(t)\right), \end{aligned} \quad (5)$$

where the $a_\uparrow \lesssim 1$ and $a_\downarrow \gtrsim 0$ characterize the state preparation (pumping) fidelity, $\epsilon_\downarrow$ ($\epsilon_\uparrow$) is the error probability that the ion is spuriously detected as being in state $|\uparrow\rangle$ ($|\downarrow\rangle$), when it actually is in state $|\downarrow\rangle$ ($|\uparrow\rangle$).

# 3 Uncertainties and corrections of the measured variables

We carefully distinguish statistical and systematic errors, and systematic shifts. A statistical error of a given quantity is denoted by $\epsilon(\cdot)$, while a systematic error is denoted by $\sigma(\cdot)$. A systematic shift is denoted by $\delta(\cdot)$, and it is accompanied by a systematic error. All shifts and uncertainties are reported as *relative* quantities. Some error sources are found to be small as compared to the dominating ones, in this case an estimation of the magnitude is given, however these error sources are not quantitatively taken into account in the total error. If this is the case, it is explicitly stated in the corresponding section. As our total error is on the $10^{-3}$ level, we do not take corrections and errors below the $10^{-4}$ level into account in the final error budget. In the cases where we report mean values $\bar{\epsilon}, \bar{\sigma}$, these quantities result from averaging over all data sets. In these cases, the individual values for each data-set rather than the indicated mean values are used for the computation of the final values.

## 3.1 Statistical Uncertainty of $\delta R$ and $b$

For each curve $i$, $R_{\pm,i,j}$ and $b$ are determined by a fit of the depolarization data to the model Eqs. 5. The uncertainties of that fit are extracted by marginalization of the data generated by the used Markov Chain Monte Carlo algorithm. It is verified by parametric bootstrapping: For fixed $R_\pm$ and $b$ values, 30 instances of simulated measurement data with binomial noise are generated and fitted in the same way as the original data, each fit yielding result values for $R_\pm$ and $b$ and their statistical errors. The standard errors determined from the actual data are in agreement with the standard deviation of the mean values found for the 30 bootstrapping datasets. Note that cross-correlations between uncertainties of $R_\pm$ and $b$ and the additional parameters introduced for characterizing SPAM errors, c.f. Sec. 2.3, are

fully taken into account due to the marginalization of the Monte Carlo data.

The individual standard error of $\delta R$ for each dataset $i,j$ is used to determine the total uncertainty of $\gamma_{PS}$, see section 4. The individual standard errors of $b$ affect the total uncertainty $\Gamma_{PD}$ and $\tau$, c.f. Eqs. 3 and 4. The means over all datasets of the relative statistical errors are

$$\begin{aligned} \bar{\epsilon}(\delta R) &= 12.1 \cdot 10^{-3} \\ \bar{\epsilon}(b) &= 24.8 \cdot 10^{-3} \end{aligned} \tag{6}$$

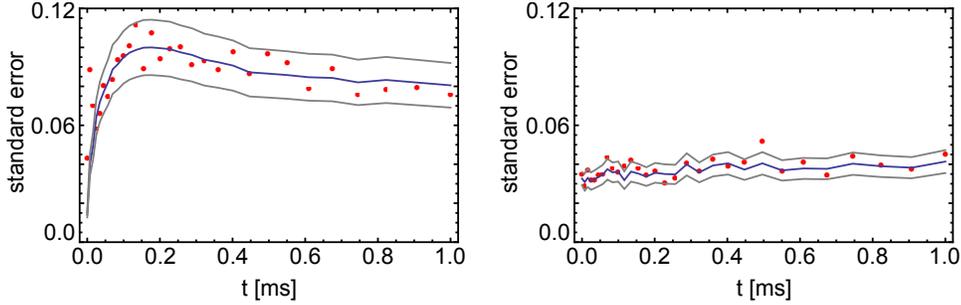

Figure 1: Comparison of expected and actual spreads of measured dark count fractions: We plot the expected population standard deviation Eq. 7 (solid black line), computed using the total estimated dark count fractions $p_\uparrow^{(\downarrow,\uparrow)}(t)$ and $N = 25$. The red points are the sample standard deviations of the dark count fractions from the 100 different acquired chunks of $N$ =25 interrogations. The solid grey lines show $2\sigma$ intervals for the expected standard sample deviations. A significant fraction of outliers would indicate additional drifts and fluctuations. The data for $p_\uparrow^{(\downarrow)}(t)$ is shown in the left panel, the data for $p_\uparrow^{(\uparrow)}(t)$ in the right panel.

The validity of the statistical errors relies on the assumption that their dominant source is the readout shot noise of the spin measurements. Possible drifts or fluctuations of the optical power or polarization during a depolarization rate measurement would corrupt this assumption. We verified that no such effects are present on a significant level in the following way: For a given data-set, for each exposure time $t$, the measurement data for $p_\uparrow^{(\uparrow,\downarrow)}(t)$ was taken in chunks comprised of $N$ =25 spin interrogations. For each chunk, coarse estimates for $p_\uparrow^{(\uparrow,\downarrow)}(t)$ are obtained by dividing the number of dark count events by the number of interrogations $N$. The sample standard deviation of these values over the set of chunks is compared to the expected population standard deviation for binomial statistics,

$$\epsilon\left(p_\uparrow^{(\uparrow,\downarrow)}(t)\right) = \sqrt{p_\uparrow^{(\uparrow,\downarrow)}(t)\left(1 - p_\uparrow^{(\uparrow,\downarrow)}(t)\right)/N}. \tag{7}$$

The results are in agreement with the assumption of binomial statistics, we can therefore exclude additional drifts and fluctuations at a significant level. An example is shown in Fig. 1.

5

## 3.2 Detuning $\Delta$

The detunings read $\Delta_j = f_j - f_{res}$, with $f_{res}$ being the so far unknown resonance frequency of the transition. The precision of the detunings $\Delta_j$ are given by the measurement precision of the absolute laser frequencies $f_j$ of the four detuning values. These are measured by a commercial wavemeter [3]. The resulting uncertainties have been quantified by determining the standard deviation of a time series of wavelength measurements of a frequency-stabilized laser, which has a linewidth orders of magnitude smaller than that measured standard deviation. These uncertainties amount to:

$$\epsilon_{wm}(\Delta_1) = 0.4 \cdot 10^{-3}$$
$$\epsilon_{wm}(\Delta_2) = 0.5 \cdot 10^{-3}$$
$$\epsilon_{wm}(\Delta_3) = 0.6 \cdot 10^{-3}$$
$$\epsilon_{wm}(\Delta_4) = 0.4 \cdot 10^{-3} \tag{8}$$

To determine the resonance frequency $f_{res}$, we note that the ratio of ac Stark shift and differential scattering rate is proportional to the detuning:

$$(-)\frac{1}{3}\frac{\Delta_{S,i,j}}{\delta R_{i,j}} \propto \Delta_j. \tag{9}$$

The minus sign holds for negative detunings. We perform a linear regression of this quantity with respect to the wavelength meter frequency $f_j$ from which the zero-crossing $f_{res}$, i.e. the resonance frequency, is obtained. The accuracies of the detunings $\Delta_j$ are limited by the accuracy of the linear regression used to determine the numerical value of $f_{res}$. As $f_{res}$ is used to compute *all* detunings, it is a systematic error. The values are:

$$\sigma_{fit}(\Delta_1) = 1.5 \cdot 10^{-3}$$
$$\sigma_{fit}(\Delta_2) = 1.8 \cdot 10^{-3}$$
$$\sigma_{fit}(\Delta_3) = 1.8 \cdot 10^{-3}$$
$$\sigma_{fit}(\Delta_4) = 1.6 \cdot 10^{-3} \tag{10}$$

## 3.3 Residual near-resonant light at 397 nm

Two laser sources provide spurious residual light weakly driving the cycling transition.

- Doppler cooling laser: The laser used for Doppler cooling and fluorescence readout is switched by an acousto-optical modulator [4] in double-pass configuration. An estimate of the residual photon scattering rate due to imperfect switch-off is obtained by initializing the qubit in $|\uparrow\rangle$ or $|\downarrow\rangle$, either before or after a 10 ms wait time before readout. This yields spurious spin flip rates of $R_-^{(res)}$ =1.1(6) Hz and $R_+^{(res)}$ =0.7(4) Hz. These rates are taken into account for two of the acquired

---
[3] WSU, High Finesse GmbH
[4] I-M110-3C10BB-3-GH27, Gooch&Housego

datasets, by modifying the model Eqs.M6, incorporating the fact that these persist throughout the entire 1 ms interval. From fitting with the modified model, obtain the following corrections:

$$\begin{aligned} \delta_{397}(\delta R) &= -4.1 \cdot 10^{-4} \\ \delta_{397}(b) &= -8.3 \cdot 10^{-4} \end{aligned} \quad (11)$$

and systematic errors

$$\begin{aligned} \sigma_{397}(\delta R) &= 4.6 \cdot 10^{-4} \\ \sigma_{397}(b) &= 6.4 \cdot 10^{-4} \end{aligned} \quad (12)$$

- Off-resonant laser: The off-resonant light near 397 nm is obtained from an amplified, frequency doubled diode laser system [5]. The amplified fundamental beam seeding the doubler has a broad background of amplified spontaneous emission (ASE), extending over about 30 nm, equivalent to $\delta_{amp} \approx 2\pi \cdot 57$ THz. The suppression with respect to the coherent light is specified to be $\chi = -40$ dB. One might argue that this background leads to additional UV light near the atomic resonance, which would effectively shorten the measured lifetime. We give a pessimistic estimate of this effect, assuming only $\chi = -30$ dB ASE suppression, perfect phase-matching for sum-frequency generation (SFG) of ASE and laser mode photons, perfect AOM diffraction even for incoherent light and that a spurious component exactly hits the atomic resonance. With a SHG cavity linewidth of $\delta_c = 2\pi \cdot 500$ kHz, we obtain a power ratio of the spurious to the main SHG component of

$$\frac{P_{spur}}{P_{SHG}} \approx \chi \frac{\delta_c}{\delta_{amp}} \approx 9 \cdot 10^{-15} \quad (13)$$

This strong suppression even under pessimistic assumptions leads us to the conclusion that this effect does not have to be taken into account. We still perform an experimental verification, where we probe depolarization after initialization in $|\downarrow\rangle$, at a fixed off-resonant pulse-time of 50 µs, corresponding to about 20% depolarization. We scan the laser frequency over a range of 1.3 GHz, which is more than one free spectral range of the SHG cavity. The scan speed is about 2 MHz/s, such that one data point integrates over about 4 MHz, which is slow enough to resolve the atomic line. No resonant features were observed in the resulting depolarization signal.

### 3.4 Residual $\pi$-polarized scattering light

The k-vector of the scattering light is aligned parallel to the magnetic field, so ideally only $\sigma$-transitions can be driven. Alignment errors of the beam might be the cause for residual $\pi$-polarized light, which is not accounted for in Eq. M4. Taking into account an the pumping to the D$_{3/2}$ caused by additional $\pi$ polarized light, the rate equations Eqs. M6 are extended, including an additional parameter $R_D$:

$$\begin{aligned} \dot{p}_\uparrow &= -R_-(1+b)\, p_\uparrow - R_D\, p_\uparrow + R_+ p_\downarrow \\ \dot{p}_\downarrow &= -R_+(1+b)\, p_\uparrow - R_D\, p_\downarrow + R_- p_\uparrow. \end{aligned} \quad (14)$$

with

$$R_D = \frac{b}{2} \frac{\epsilon_0^2}{\epsilon_+^2} R_+. \quad (15)$$

---

[5]TA SHGpro, Toptica Photonics AG

The solutions of the rate equations are of the same form as for Eq. M6, however with altered coefficients

$$\bar{R} = (1+b)(R_- + R_+) + R_D \qquad (16)$$
$$\tilde{R}^2 = \bar{R}^2 - 4(2+b)(R_D + bR_-)R_+. \qquad (17)$$

If the data is now fitted using this modified model, we obtain values of about $R_D = 3(3)$ Hz, and altered values for $\delta R, b$. The mean value of the deviations of $\delta R, b$ are used to compute the corrections:

$$\bar{\delta}_\pi(\delta R) = 3.3 \cdot 10^{-3}$$
$$\bar{\delta}_\pi(b) = -2.4 \cdot 10^{-3}, \qquad (18)$$

and the spread of the deviations within each data-set determines the systematic error

$$\bar{\sigma}_\pi(\delta R) = 2.9 \cdot 10^{-3}$$
$$\bar{\sigma}_\pi(b) = 2.6 \cdot 10^{-3} \qquad (19)$$

## 3.5 Differential Stark Shift $\Delta_S$

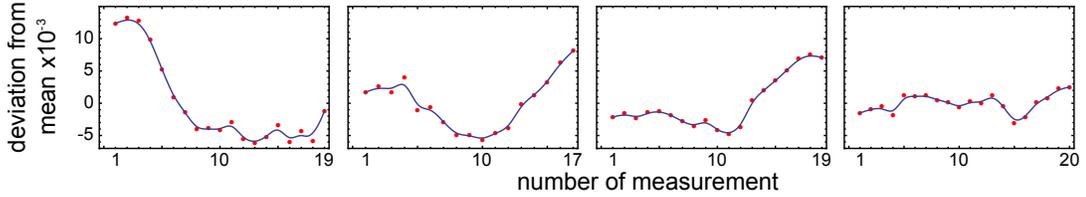

Figure 2: The red points depict the relative deviations of the ac Stark shift values $\Delta_{S,i,j}$ from their mean for all four measurement sessions. The blue curves are the B-spline functions defined by the points. The RMS deviation of the points from the spline is used as the statistical error.

We carried out the bootstrapping method from Sec. 3.1 for determining the uncertainty of the ac Stark Shift $\Delta_S$. The result is that the uncertainty from the binomial readout statistics of the measurement would lead to insignificant uncertainties on the $10^{-7}$ level. Systematic effects, i.e. temporal drifts, are clearly dominant.
$\Delta_{S,i,j}$ is for each dataset $i,j$ is determined by the mean of two measurements, taken before and after the scattering rate measurement yielding $\delta R_{i,j}$. The drift of the ac Stark shift drift over the measurement time each detuning $\Delta_i$ is slow and therefore mostly resolved, thus the assumption of a linear drift between two consecutive stark shift measurements is justified. To account for residual drifts on a timescale which is not resolved, we use the measured $\Delta_S$ values as control points to define a B-spline function as shown in Fig. 2. The RMS of the difference of the measurement result and the corresponding B-spline value is treated as the overall inaccuracy $\epsilon(\Delta_{S,i,j})$ of all $\Delta_S$ values pertaining to detuning $\Delta_i$. The mean value of those relative uncertainties reads

$$\bar{\epsilon}(\Delta_S) = 3.0 \cdot 10^{-4} \qquad (20)$$

### 3.6 Zeeman shift

We do not take the Zeeman shifts of the atomic levels ($\Delta_Z = 2\pi \cdot 13.34$ MHz splitting between $|\uparrow\rangle$ and $|\downarrow\rangle$) into account. However, as the atomic resonance frequency is determined via the ac Stark shift caused by the predominantly $\sigma_+$ polarized beam, the detuning $\Delta$ specifically refers to the $|S_{1/2}, m_J = -1/2\rangle \leftrightarrow |P_{1/2}, m_J = +1/2\rangle$ transition *including* its Zeeman shift. Corrections are only necessary for the transitions driven by the weak $\sigma_-$ and $\pi$ polarization components. The corresponding systematic errors are estimated to be

$$\begin{aligned}
\sigma_Z(\delta R) &\approx \frac{\epsilon_{0,-}^2}{\epsilon_+^2} \frac{\Delta_Z}{\Delta} \approx \frac{R_-}{R_+} \frac{\Delta_Z}{\Delta} \lesssim 3 \cdot 10^{-5} \\
\sigma_Z(\Delta_S) &\lesssim 2 \cdot 10^{-5}
\end{aligned} \quad (21)$$

### 3.7 Influence of Micromotion

Both the scattering rates $R_+$, $R_-$ and the Stark shift $\Delta_S$ are altered in the presence of micromotion along the propagation direction of the off-resonant beam. The micromotion causes effective frequency modulation (FM), characterized by the trap drive frequency $\Omega_{rf} = 2\pi \cdot 25$ MHz and the modulation index $\beta$. We obtain

$$R_\pm^{(mm)} \approx \gamma_{PS} \frac{\epsilon_\pm^2 + \epsilon_0^2}{9} \sum_{n=-\infty}^{+\infty} J_n^2(\beta) \frac{\Omega^2}{4(\Delta + n\Omega_{rf})^2} \quad (22)$$

$$\Delta_S^{(mm)} \approx \frac{\epsilon_+^2 - \epsilon_-^2}{3} \sum_{n=-\infty}^{+\infty} J_n^2(\beta) \frac{\Omega^2}{4(\Delta + n\Omega_{rf})} \quad (23)$$

By measuring the the shape of the resonance fluorescence line by recording fluorescence photons at low power and across resonance, an upper bound for the modulation Index of $\beta = 3.5$ is obtained. Note that the micromotion-induced frequency modulation leads to symmetric FM sidebands, therefore the effects cancel out to first order. Due to the $\Delta^{-1}(\Delta^{-2})$ scaling behavior of the ac Stark shift (scattering rates), *both* quantities are slightly increased by FM components on the side towards the resonance, such that the total effect partially cancels out. Thus, we give an estimate for the systematic error of the resulting $\gamma_{PS}$ rather than for $\Delta_S$ and $\delta R$. Using realistic values of $\Omega^2, \epsilon_+^2, \epsilon_-^2$ and $\Delta$, we obtain

$$\sigma_{mm}(\gamma_{PS}) = \frac{1}{\delta R}\left(\delta R^{(mm)} - \delta R\right) - \frac{1}{\Delta_S}\left(\Delta_S^{(mm)} - \Delta_S\right) \approx 1.0 \cdot 10^{-4} \quad (24)$$

### 3.8 Influence of near-resonance-effects

The expressions for $\delta R$ Eq. M2 and $\Delta_S$ Eq. M1 are only valid if $|\Delta| \gg \gamma_{PS}, \Omega$. Our experimental conditions are $\gamma_{PS}/|\Delta| \lesssim 2 \cdot 10^{-3}, \Omega/|\Delta| \lesssim 4 \cdot 10^{-2}$, we therefore investigate how the approximations affect the results.

The *measured*, i.e. actual scattering rate $\delta R^{(meas)}$ is *smaller* than the far-off-resonance approximation

Eq. M2. Including power broadening and finite natural linewidth, the scattering rate is given by

$$\begin{aligned}
\delta R^{(meas)} &= \gamma_{PS}\frac{\epsilon_+^2 - \epsilon_-^2}{9}\frac{\Omega^2}{4\Delta^2 + \frac{2}{3}\Omega^2 + \gamma_{PS}^2} \\
&\approx \gamma_{PS}\frac{\epsilon_+^2 - \epsilon_-^2}{9}\Omega^2\left(\frac{1}{4\Delta^2} - \frac{\frac{2}{3}\Omega^2 + \gamma_{PS}^2}{16\Delta^4}\right) \\
&= \delta R + \delta R'
\end{aligned} \quad (25)$$

We thus have

$$\delta R' = -\delta R\frac{\frac{2}{3}\Omega^2 + \gamma_{PS}^2}{4\Delta^2} \quad (26)$$

The correction is obtained by invoking the expression Eq. M3 for the decay rate $\gamma_{PS}$:

$$\begin{aligned}
\gamma_{PS} &= \frac{3\Delta}{\Delta_S}(\delta R^{(meas)} - \delta R') \\
&= \frac{3\Delta}{\Delta_S}\delta R^{(meas)}\left(1 + \delta_{broad}(\delta R)\right)
\end{aligned} \quad (27)$$

such that

$$\delta_{broad}(\delta R) = -\frac{\delta R'}{\delta R^{(meas)}} \approx 2\Delta_S/\Delta \quad (28)$$

We obtain

$$\begin{aligned}
\delta_{broad}(\delta R_1) &= 2.0 \cdot 10^{-4} \\
\delta_{broad}(\delta R_2) &= 2.6 \cdot 10^{-4} \\
\delta_{broad}(\delta R_3) &= 2.9 \cdot 10^{-4} \\
\delta_{broad}(\delta R_4) &= 2.1 \cdot 10^{-4}.
\end{aligned} \quad (29)$$

For $\Delta_S$, no expression corresponding to Eq. 25 is available, we therefore extract values from a numerical simulation using a three-level generalized optical Bloch equation. For $\delta_{broad}(\Delta_S) = (\Delta_S^{(meas)} - \Delta_S)/\Delta_S^{(meas)}$, we obtain

$$\begin{aligned}
\delta_{broad}(\Delta_{S,1}) &= 1.0 \cdot 10^{-4} \\
\delta_{broad}(\Delta_{S,2}) &= 1.4 \cdot 10^{-4} \\
\delta_{broad}(\Delta_{S,3}) &= 1.4 \cdot 10^{-4} \\
\delta_{broad}(\Delta_{S,4}) &= 1.0 \cdot 10^{-4}.
\end{aligned} \quad (30)$$

### 3.9 Finite $D_{3/2}$-lifetime and residual 866 nm repump light

In the model Eqs. M6, the lifetime of the $D_{3/2}$-state is considered to be infinite. The most recent experimental value is 1176(11) ms (Kreuter et. al., PRA 71, 2005). Residual light near 866 nm also depletes population accumulated in the $D_{3/2}$ state at a rate $R_{dep}$, effectively reducing its lifetime. We measure the residual power in the imperfectly switched off beam, using a fibre-coupled avalanche

photodiode, to about 6 pW. In the switched-on state, the optical power is 86 μW. For observing the amount of resonance fluorescence versus optical power at 866 nm, we infer a saturation parameter of 23(5). This leads to a saturation parameter in the switched-off state of $S_{off}^{(866)}$ =1.6(8)·10$^{-6}$. The depletion rate of the $D_{3/2}$ state is then

$$R_{D3/2}^{(dep)} = \frac{\gamma_{PS}}{2} \frac{S_{off}^{(866)}}{1+4\bar{\delta}^2/\gamma_{PD}^2}. \quad (31)$$

Here, $\bar{\delta}$ is the average detuning of the sub-transitions between the $P_{1/2}$ and $D_{3/2}$ manifolds, considering the Zeeman splitting (see Sec. 3.6) and the repump laser to be tuned onto the blue side of the entire manifold. We obtain a laser induced depletion rate of 1.2(1.0) Hz. Thus, the $D_{3/2}$ state is depleted at a total rate of about $R_{D3/2}^{(dep)}$ =2.2(1.0) Hz.

For estimating the impact of the depletion on the result for $\delta R$, we modify the rate equations Eqs. M6 to be

$$\begin{aligned}
\dot{p}_\uparrow &= -R_-(1+b)\,p_\uparrow + R_+ p_\downarrow + \tfrac{1}{2} R_{D3/2}^{(dep)}(1-p_\uparrow - p_\downarrow) \\
\dot{p}_\downarrow &= -R_+(1+b)\,p_\downarrow + R_- p_\uparrow + \tfrac{1}{2} R_{D3/2}^{(dep)}(1-p_\uparrow - p_\downarrow).
\end{aligned} \quad (32)$$

Using these for example data in our fit procedure, we infer the systematic shifts

$$\begin{aligned}
\delta_{\gamma D3/2}(\delta R) &= 3.2 \cdot 10^{-4} \\
\delta_{\gamma D3/2}(b) &= 1.6 \cdot 10^{-3}
\end{aligned} \quad (33)$$

and systematic errors

$$\begin{aligned}
\sigma_{\gamma D3/2}(\delta R) &= 1.6 \cdot 10^{-4} \\
\sigma_{\gamma D3/2}(b) &= 7 \cdot 10^{-4}
\end{aligned} \quad (34)$$

### 3.10 Influence of the $P_{3/2}$ state

The $P_{3/2}$ is separated from the $P_{1/2}$ manifold by the fine-structure-splitting of $\Delta_{FS} \approx 2\pi \cdot 6.69$ THz. The coupling strength $\Omega^2$ is increased with respect to the $P_{1/2}$ state by a factor of 2, given by the ration of the squared reduced dipole matrix elements, which we indicate explicitly. The presence of these levels affects both the Stark-shift and the scattering rates. For the scattering, we approximate the decay rate to be the same as for the $^2P_{1/2}$ state, and we take only the predominant $\sigma_+$ polarization component into account, and we neglect decay into the $D_{3/2}$ and $D_{5/2}$ states. The additional scattering rate then reads:

$$R_+^{(P3/2)} \approx \gamma_{PS} \frac{2}{3} \frac{\epsilon_+^2}{6} \frac{2\Omega^2}{4(\Delta_{FS}-\Delta)^2} \approx \gamma_{PS} \frac{1}{9} \frac{2\Omega^2}{4\Delta_{FS}^2} \quad (35)$$

Note that the for the $P_{3/2}$ state, due to the different Clebsch-Gordan coefficients, the branching ratio of decays with spin-flip to decay without spin-flip is reversed with respect to the $P_{1/2}$ state, however the scattering rate has the same prefactor as in Eq. M2. The excitation channel $S_{1/2}, m_J = +1/2 \to P_{3/2}, m_J = +3/2$ contributes only to elastic scattering. Interference effects between different pathways

11

do not play a role as the emitted photons are distinguishable either by wavelength or by polarization/emission direction. The correction to the ac Stark shift reads:

$$\Delta_S^{(P3/2)} \approx \left(\frac{\epsilon_+^2}{2} - \frac{\epsilon_+^2}{6}\right)\frac{2\Omega^2}{4(\Delta_{FS} - \Delta)} \approx \frac{\epsilon_+^2}{3}\frac{2\Omega^2}{4\Delta_{FS}} \quad (36)$$

For the $P_{3/2}$ state, the transition $S_{1/2}, m_J = +1/2 \to P_{3/2}, m_J = +3/2$ contributes to the differential ac Stark shift and has a larger Clebsch-Gordan coefficient than the transition $S_{1/2}, m_J = -1/2 \to P_{3/2}, m_J = +1/2$. The reversed sign has to be accounted for as we only measure absolute values of $\Delta_S$. For $\Delta > 0$, the magnitude of $\Delta_S$ is increased, for $\Delta < 0$ it is decreased.
We finally obtain

$$\delta_{P3/2}(\Delta_{S,i,j}) = \frac{\Delta_i}{3\Delta_{FS}} \approx 1.0 \cdot 10^{-3}\frac{\Delta_i}{2\pi \cdot 10 \text{ GHz}} \quad (37)$$

$$\delta_{P3/2}(\delta R_{S,i,j}) = \frac{\Delta_i^2}{3\Delta_{FS}^2} \approx 1.5 \cdot 10^{-6}\left(\frac{\Delta_i}{2\pi \cdot 10 \text{ GHz}}\right)^2 \quad (38)$$

Note that the $\delta_{P3/2}(\Delta_{S,i,j})$ depend on the sign of $\Delta_i$. We do not include systematic errors due to the small magnitudes of these corrections.

### 3.11 Fit bias

Parameter estimation from measurement data might be affected by a systematic bias, depending on the used estimator. We generated artificial measurement data from the model Eq. M6 with fixed values of $R_\pm, b$ with binomial errors. We find that for increasing number of measurements, i.e. vanishing noise, the fit results for $R_\pm, b$ converges to the preset values. We thus do not take any bias into account in our error budget.

### 3.12 SPAM errors

For the determination of the differential scattering rate $\delta R$, SPAM errors are taken into account by employing the model introduced in Sec. 2.3 for the parameter estimation. The uncertainty of the additional parameters and their cross-correlation with $R_\pm, b$ is already taken into account, as the statistical error $\epsilon(\delta R)$ is obtained by marginalizing over the sample data from the Monte Carlo parameter estimation algorithm.

## 4 Total uncertainties and corrections of $\gamma_{PS}, \gamma_{PD}$ and $\tau$

The uncorrected values for $\gamma_{PS}$ for each detuning $\Delta_i$ are obtained from averaging:

$$\gamma'_{PS,i} = \frac{1}{N_i}\sum_{j=1}^{N_i} 3\Delta_i \frac{\delta R_{i,j}}{\Delta_{S,i,j}}$$

$$b'_i = \frac{1}{N_i}\sum_{j=1}^{N_i} b_{i,j} \quad (39)$$

The corrected values are obtained by adding all relative corrections:

$$\begin{aligned}
\gamma_{PS,i}/\gamma'_{PS,i} &= 1 + \delta_{397}(\delta R) + \delta_{broad}(\delta R_i) + \delta_{\gamma D3/2}(\delta R) + \delta_{\pi}(\delta R) \\
&\quad - \delta_{broad}(\Delta_{S,i}) - \delta_{P3/2}(\Delta_{S,i}) \\
b_i/b'_i &= 1 + \delta_{397}(b) + \delta_{\gamma D3/2}(b) + \delta_{\pi}(b)
\end{aligned} \quad (40)$$

The statistical errors are obtained from error propagation:

$$\begin{aligned}
\epsilon(\gamma_{PS,i,j})^2 &= \epsilon(\Delta_i)^2 + \epsilon(\Delta_{S,i,j})^2 + \epsilon(\delta R_{i,j})^2 \\
\epsilon(b_i)^2 &= \frac{1}{N_i^2} \sum_{j=1}^{N_i} \epsilon(b_{i,j})^2
\end{aligned} \quad (41)$$

For taking the systematic errors into account, we face the problem that most of their correlations are not quantitatively available. However, it can be reasoned that they are due to well-distinguished physical effects and thus influence the resulting quantities in ways which are hardly related to each other. [6]. According to JCGM guidelines [7] we therefore neglect possible correlations and add them in quadrature:

$$\begin{aligned}
\sigma(\gamma_{PS,i})^2 &= \sigma_{mm}(\gamma_{PS})^2 + \sigma_{fit}(\Delta_i)^2 + \sigma_{397}(\delta R)^2 + \sigma_{\gamma D3/2}(\delta R)^2 + \sigma_{\pi}(\delta R_i)^2 \\
\sigma(b_i)^2 &= \sigma_{397}(b)^2 + \sigma_{\gamma D3/2}(b)^2 + \sigma_{\pi}(b_i)^2
\end{aligned} \quad (42)$$

The final values are obtained by a weighted average over the four results:

$$\begin{aligned}
\gamma_{PS} &= \frac{1}{\sum_i N_i} \sum_i N_i \gamma_{PS,i} \\
b &= \frac{1}{\sum_i N_i} \sum_i N_i b'_i
\end{aligned} \quad (43)$$

In Table I in the main manuscript, the mean values of all statistical and systematic uncertainties of $\gamma_{PS}$ are given. When added in quadrature, they yield the total uncertainty. In the following calculation, we pursue a more detailed approach, where the uncertainties of the 65 measurement runs are evaluated individually, before the mean is taken. The difference of the results however turns out to be insignificant ($< 1 \times 10^{-4}$).

The total statistical errors are given by

$$\begin{aligned}
\epsilon(\gamma_{PS})^2 &= \frac{1}{\left(\sum_i N_i\right)^2} \sum_{i,j} \epsilon(\gamma_{PS,i,j})^2 \\
\epsilon(b)^2 &= \frac{1}{\left(\sum_i N_i\right)^2} \sum_i N_i^2 \epsilon(b_i)^2.
\end{aligned} \quad (44)$$

---

[6] Note that all systematic errors depend on $\Delta$, except for $\sigma_{fit}(\Delta_i)$ - the error of $\Delta$ itself. The resulting corrections from these correlations are on the $10^{-6}$ level and are thus insignificant.

[7] Joint Committee for Guides in Metrology, *Evaluation of measurement data - Guide to the expression of uncertainty in measurement*, p. 62 (2008), http://www.bipm.org/en/publications/guides/gum.html.

For the total systematic errors, we assume complete correlation between experiments at different detunings and average them linearly:

$$\begin{aligned}
\sigma(\gamma_{PS}) &= \frac{1}{\sum_i N_i} \sum_i N_i \sigma(\gamma_{PS,i}) \\
\sigma(b) &= \frac{1}{\sum_i N_i} \sum_i N_i \sigma(b_i),
\end{aligned} \quad (45)$$

The total measurement uncertainties result from adding statistical and systematic errors in quadrature:

$$\begin{aligned}
\mathcal{E}(\gamma_{PS})^2 &= \epsilon(\gamma_{PS})^2 + \sigma(\gamma_{PS})^2 \\
\mathcal{E}(b)^2 &= \epsilon(b)^2 + \sigma(b)^2
\end{aligned} \quad (46)$$

For the radiative decay rate to the $D_{3/2}$ state, $\gamma_{PD} = (b/3)\gamma_{PS}$, we thus find

$$\mathcal{E}(\gamma_{PD})^2 = \epsilon(b)^2 + \epsilon(\gamma_{PS})^2 + \sigma(b)^2 + \sigma(\gamma_{PS})^2 \quad (47)$$

and correspondingly for the total lifetime $\tau = 1/(\gamma_{PS} + \gamma_{PD})$:

$$\mathcal{E}(\tau)^2 = \epsilon(\gamma_{PS})^2 + \sigma(\gamma_{PS})^2 + \frac{b^2}{(3+b)^2}\left(\epsilon(b)^2 + \sigma(b)^2\right). \quad (48)$$



\end{document}